# Optimal quantum control using randomized benchmarking

J. Kelly,[1,*] R. Barends,[1,*] B. Campbell,[1] Y. Chen,[1] Z. Chen,[1] B. Chiaro,[1] A. Dunsworth,[1] A. G. Fowler,[1,2] I. Hoi,[1] E. Jeffrey,[1] A. Megrant,[1] J. Mutus,[1] C. Neill,[1] P.J.J. O'Malley,[1] C. Quintana,[1] P. Roushan,[1] D. Sank,[1] A. Vainsencher,[1] J. Wenner,[1] T. C. White,[1] A. N. Cleland,[1] and John M. Martinis[1]

[1]*Department of Physics, University of California, Santa Barbara, CA 93106, USA*
[2]*Centre for Quantum Computation and Communication Technology,
School of Physics, The University of Melbourne, Victoria 3010, Australia*

We present a method for optimizing quantum control in experimental systems, using a subset of randomized benchmarking measurements to rapidly infer error. This is demonstrated to improve single- and two-qubit gates, minimize gate bleedthrough, where a gate mechanism can cause errors on subsequent gates, and identify control crosstalk in superconducting qubits. This method is able to correct parameters to where control errors no longer dominate, and is suitable for automated and closed-loop optimization of experimental systems.

Quantum control is inherently analog [1], so gate control parameters have to be set precisely to enable fault-tolerant quantum computing [2, 3]. With gate fidelities approaching the fault-tolerant threshold [4–6], characterizing and reducing the remnant errors become increasingly challenging. Quantum process tomography can completely characterize a gate, decomposing a process into Pauli or Kraus operators [7, 8]. However, improving gates is complicated: gate parameters map non-intuitively onto the process matrix, and state preparation and measurement errors (SPAM) can be confused with process errors.

Here, we present a different approach to achieve high fidelity gates. We use Clifford-based randomized benchmarking (RB) [9, 10] to map gate errors onto control parameters. We show how the data can be directly fed back to optimize gates. The method is fast and scales to arbitrary precision as the sensitivity to fractional error is independent of gate fidelity. We experimentally show that it works well for several real problems. We apply it to general quantum control problems, such as gate optimization, gate bleedthrough [11] and crosstalk. In particular we demonstrate closed-loop optimization with nonorthogonal parameters in a real, noisy quantum system. As RB is platform-independent, our approach is in principle applicable to a variety of quantum systems.

In standard RB, gate are characterized by measuring the fidelities of sequences with varying lengths. We experimentally show that optimizing the *sequence* fidelity at fixed length improves the *gate* fidelity. We call this approach optimized randomized benchmarking for immediate tune-up (ORBIT).

As a testbed, we use a five qubit superconducting system based on the Xmon transmon design [12]. Here, XY control is achieved with microwave pulses and Z control with DC current pulses which modulate the qubit frequency. Qubits are coupled capacitively. Qubit frequencies lie between $f_{10} = 4$ and 6 GHz, and qubit nonlinearities $\Delta/2\pi$ are around 220 MHz. This device is an ideal platform for optimizing for small errors, as we have obtained high fidelity single- and two-qubit gates (see Ref. [5]).

We start with a simple test case where we optimize a single-qubit 90 degree rotation about the X-axis in the Bloch sphere representation (X/2 gate). This gate is implemented by a mi-crowave pulse with a cosine envelope (Fig. 1a inset) centered around frequency $f$ with amplitude $A$. As the Xmon transmon qubit is a multilevel system, we must apply a quadrature correction term with coefficient $\alpha$ to minimize leakage to higher levels [13, 14]. First, we determine the gate fidelity using RB, then measure how control errors affect fidelity of sequences.

In Clifford-based RB, random Clifford rotations are inserted between the gate under test to ensure that it is applied to a representative set of states. The single-qubit Clifford gates are the group of rotations that preserve the two polar and four equally spaced equator states on the Bloch sphere, and are able to generate a representative set of states to remove bias from gate error. To quantify the X/2 fidelity, we first measure a reference curve by applying many sequences of ran-dom Cliffords, appended by recovery Cliffords $C_r$ that make the ideal operation the identity. As we initialize the qubit in the ground state, the ground state population becomes the sequence fidelity. Randomization makes the sequence fidelity follow an exponential decay from the accumulation of gate-aspecific errors as $Ap_{\mathrm{ref}}^m + B$, with gate errors captured in the characteristic scale $p_{\mathrm{ref}}$ ( see Fig. 1a). The SPAM errors affect $A$ and $B$, but not the rate of the decay. Individual gate fidelities are evaluated by interleaving a specific gate between Cliffords, generating a decay curve with scale $p_{\mathrm{gate}}$. By subtracting away the reference curve, we get the gate error $r_{\mathrm{gate}} = (1 - p_{\mathrm{gate}}/p_{\mathrm{ref}})(d-1)/d$ [15], with $d = 2^n$ a function of the number of qubits $n$; here $n = 1$. Each point in $m$ is an average of the fidelity of $k$ different random sequences. We find the fidelity of this X/2 gate to be 0.9995 ($k = 40$).

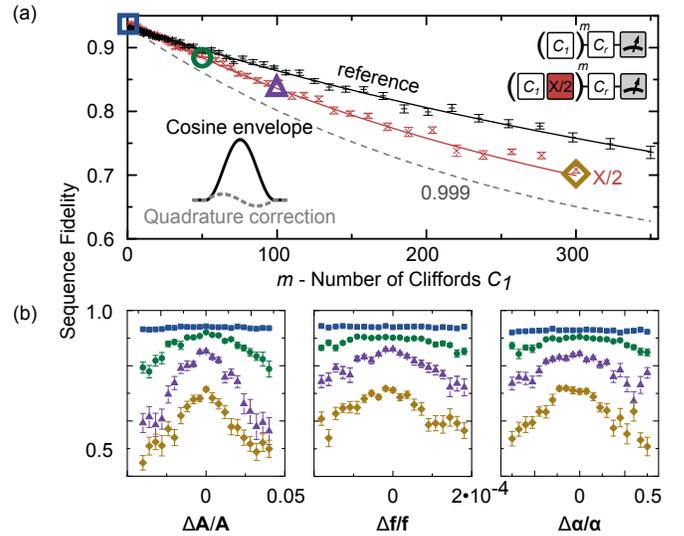

FIG. 1. (Color online) (a) Single qubit randomized benchmarking ($k = 40$). The reference experiment uses sequences of random Cliffords only (black hexagons); a X/2 gate is tested by interleaving it with random Cliffords (red stars). The X/2 gate has a cosine envelope pulse shape and quadrature correction (inset). (b) Sequence fidelities versus parameters: the pulse amplitude $A$, frequency $f$, and coefficient $\alpha$ ($k = 20$). The sequences are measured at $m = 1$ (square), 50 (circle), 100 (triangle), and 300 (diamond).

For the data in Fig. 1b, we set $m = 1, 50, 100, 300$ and measure the sequence fidelity as we vary each of the gate parameters from their optimum. As expected, we find that longer length sequences drop more rapidly in fidelity away from the maximum, indicating an increased sensitivity to gate error with sequence length. It is this feature that opens a viable route to optimizing arbitrarily high fidelity gates: sensitivity can be maintained by doubling $m$ when the error is halved [16].

In the rest of this Letter, we demonstrate that ORBIT is applicable to a variety of non-trivial parameterized tune-up problems, such as entangling gate optimization with non-orthogonal parameters, improving waveform control for reducing gate bleedthrough, and minimizing crosstalk in a multi-qubit system. We emphasize that these applications are issues of prime importance to high fidelity and scaling up to larger qubit systems [17].

We start by applying ORBIT to a controlled-phase (CZ) entangling gate with two qubits, as described in [5, 18]. With the addition of many non-orthogonal gate parameters and larger Hilbert space, this is a significant increase in complexity compared to the X/2 gate. The CZ gate is performed by moving a qubit along an adiabatic trajectory in frequency [18] (see inset Fig. 2a) which brings the $|11\rangle$ and $|02\rangle$ avoided level crossing near resonance, generating a conditional phase. The fidelity of this gate is sensitive to the frequency trajectory, as deviations from the ideal can cause non-adiabatic leakage errors to $|02\rangle$. The gate depends on eight parameters that follow straightforwardly from theory (see Ref. [18]).

The direct mapping that ORBIT provides between the control parameters and gate fidelity allows for automated optimization. Here, we used the Nelder-Mead algorithm for closed-loop control. As a metric, we use sequences ($m = 30$) composed of gates from the two-qubit Clifford group $C_2$, generated with an average of 1.5 CZ gates per Clifford [5]. These CZ gates dominate the error, making the reference fidelity a metric for CZ gate fidelity. Figure 2a shows the reference curves before (blue squares) and after (red circles) optimization. The average error per Clifford was reduced from $r = 0.0361$ to $r = 0.0188$, consistent with an improvement in gate fidelity from 0.984 to 0.993 (see Ref. [16] for interleaved data). Figure 2b shows the evolution of the sequence fidelity versus number of evaluations, starting with the blue square; it initially varies strongly with small parameter changes, underlining the sensitivity of this method, and eventually converges on optimal parameters (red circle). The inset of Fig. 2a shows the small change in waveform shape (5 MHz in magnitude) that improves fidelity.

Figure 2 illustrates the advantages of this approach. First, we can identify and remedy small errors in an environment with noise; we optimize parameters to where gate errors are no longer dominated by control imperfections (see Ref. [5] for a representative error budget for a similar experiment). Second, our approach is fast: the total number of measurements is 18000 ($k = 20$ sequences, 900 repetitions each), which can be performed in 2 seconds with our system. Third, the optimization is model-free, which is a powerful tool as the system Hamiltonian is not always known to high precision. We be-lieve this will be critical to improving gates beyond current fidelities.

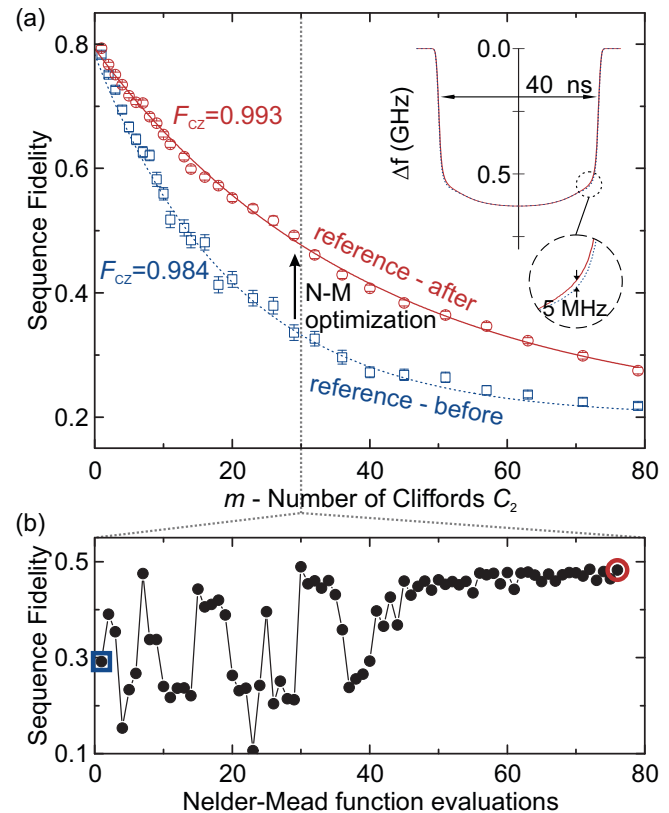

FIG. 2. (Color online). Optimizing the fidelity of a two-qubit CZ gate. (inset a) One qubit undergoes an effective adiabatic trajectory in frequency that brings the $|11\rangle$ and $|02\rangle$ near resonance, producing a conditional phase. (a) The sequence fidelity of the reference curve versus number of two-qubit Cliffords before (blue squares) and after (red circles) optimization ($k = 50$). This optimization has shifted the shoulder of the trajectory by 5 MHz (inset). (b) The change in sequence fidelity at $m = 30$ versus Nelder-Mead function evaluations ($k = 20$), starting at a fidelity of 0.3 (blue square), and converging on a sequence fidelity of 0.5 (red circle). The fidelity of the CZ improved from $F_{\mathrm{CZ}} = 0.984$ to $F_{\mathrm{CZ}} = 0.993$, measured using interleaved RB [16].

We have used the Nelder-Mead algorithm for automated tune-up as it is a gradient-free method, and therefore less sensitive to noise. While ORBIT was used for the "last mile" – where gate parameters are initially near the global optimum – algorithms that search for a global optimum could in principle also be used. We therefore envision possible applications in experimentally implementing model-free gates. One example is numerical optimal control [19, 20], where pulses are discretized into pixels. This technique can be used on the full Hamiltonian without approximation, can optimize for robustness against noise or experimental parameters, and can generate gates as fast as the "quantum speed limit" [21]. Typically, these gates are computed to machine precision assuming an idealized Hamiltonian. However, experimentally implementing such gates is hindered by differences between the assumed and actual system Hamiltonians. ORBIT could provide a bridge for direct, high-precision optimization of such gates to arbitrary fidelity in an experimental system, such as outlined in the Ad-HOC approach (see Ref. [22]).

We now use ORBIT to minimize gate bleedthrough; this is a particularly harmful problem because it causes gate-specific errors on potentially many subsequent gates. Gate bleedthrough occurs when the mechanism for implementing a gate is not adequately turned off at the end. Physical mechanisms include reflections of control pulses, stray inductance in control lines, and amplifier slew rates for microwave systems. Gate bleedthrough is challenging to characterize and correct, because the entire time domain response must be optimized. Here, ORBIT has a distinct advantage by capturing all gate bleedthrough errors in the sequence fidelity.

We reduce gate bleedthrough from a detune operation which is implemented using a square step pulse on the qubit frequency control line. The qubit is detuned for 35 ns by -0.37 GHz, acquiring a single-qubit phase $\phi = 13 \cdot 2\pi$. These current pulses can detune the qubit during subsequent gates if not properly leveled, as illustrated in the top inset of Fig. 3a. In the bottom inset, we measure deviations $\delta\phi(t)$ from the ideal acquired qubit phase before and after correction. We compensate the waveform for stray inductances and reflections in the line by applying an inverse transfer function with two poles, expressed in terms of the step response: $\Theta'(t) = \Theta(t)[1 + \sum_i a_i \exp(-\gamma_i t)]$, with $\Theta(t)$ the Heaviside step function, and amplitudes $a_i$ and rates $\gamma_i$. In Fig. 3a, the error of a Clifford plus step pulse is reduced from $r = 0.011$ to $r = 0.003$ by Nelder-Mead optimization. The sequence fidelity and evolution of the parameters $a_i$, $\gamma_i$, and accumulated qubit phase are shown to converge in Fig. 3b and Fig. 3c. Gate bleedthrough is reduced as evidenced in the improved sequence fidelity. Additionally, the remnant qubit phase $\delta\phi(t)$ is markedly flatter after the detuning pulse (see the bottom inset of Fig. 3a). This demonstrates that gate bleedthrough can be minimized without the need for a full time-domain characterization.

We also apply ORBIT to optimization problems relevant to large systems. One of the greatest challenges in scaling up to larger quantum systems is to maintain addressability over single qubits. A major obstacle is control crosstalk, where control pulses for one qubit affect others. In our architecture, we minimize control crosstalk between nearest neighbors by alternating the qubit frequency; next-nearest neighbors however are prone to crosstalk due to the smaller frequency dif-ference (see Fig. reffig:xtalka). One of the difficulties in minimizing crosstalk lies in characterizing its effect on gate fidelity. Here, ORBIT provides an elegant solution by directly mapping the gate error onto the relevant parameters, through the comparison of isolated and simultaneous application of random single-qubit Clifford sequences [23].

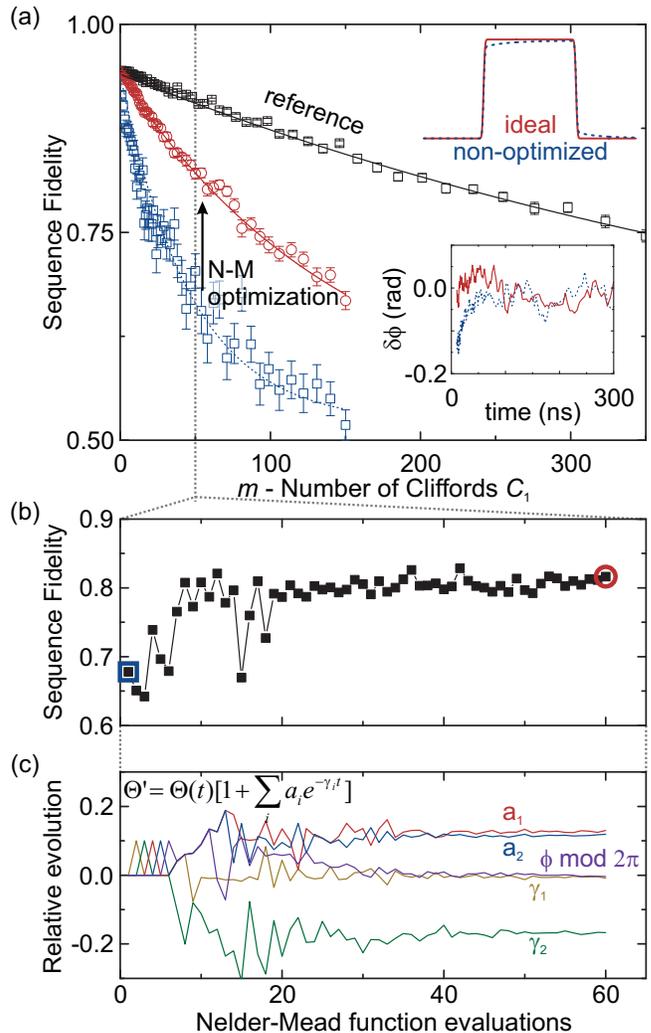

FIG. 3. (Color online) Reducing gate bleedthrough. (top inset a) For rotating the state around the Z axis (Bloch sphere representation), the qubit frequency is detuned by a step pulse, with $t_{\text{gate}} = 35$ ns and large frequency change $\Delta f = -0.37$ GHz which is ideally flat (solid red). Non-idealities in control and wiring bring about a non-trivial deformation of the waveform (dashed blue), causing gate bleedthrough. (a) Sequence fidelity vs number of Cliffords for the reference (black squares), and interleaved with the step pulse. The error per Clifford plus step pulse is reduced using ORBIT from $r = 0.011$ (blue squares) to $r = 0.003$ (red circles). With this improvement, the remnant qubit phase $\delta\phi(t)$ after the step pulse is notably more constant (bottom inset), determined via quantum state tomography ($k = 40$). (b) Sequence fidelity during the Nelder-Mead algorithm ($k = 30$). (c) Evolution of transfer function parameters, written in terms of the step response $\Theta'$.

We start by measuring the reference fidelity curve for qubit labeled $Q_2$, shown in Fig. 4b. From the decay, we find an average error per Clifford of $r_c = 0.001$, consistent with the average single qubit gate fidelity of $F = 0.9995$. The colored regions indicate different ranges in reference fidelity; we use this as a map to infer the gate fidelity from the sequence fidelity. Next, we monitor the sequence fidelity (with $m = 35$) of $Q_2$ while sending pulses for single-qubit Cliffords down the control line of $Q_0$. We can ignore the state of $Q_0$. We vary both the detuning $\delta$ and gate length $t_\text{gate}$ for pulses on the $Q_0$ line, while keeping the product of gate length and amplitude fixed to mimic control crosstalk. The inferred gate fidelity of qubit $Q_2$ is shown in Fig. 4c. The red regions indicate minimal added error from crosstalk ($< 0.05\%$), while the blue regions show significant increase in error ($> 1\%$). Clear signatures of infidelity appear when crosstalk signals are resonant with the qubit transition frequencies $f_{10}$ or $f_{21}$, as illustrated in blue in Fig. 4a, and fall off with detuning and gate length as expected.

The data in Fig. 4 demonstrates that ORBIT can provide a map to visualize and optimize control crosstalk in a straightforward manner, without the need to characterize or recalibrate the pulses on qubit $Q_0$. This technique could in principle also be used for crosstalk reduction methods that reduce spectral power at overlapping frequencies (see Ref. [24]).

In using ORBIT, we explicitly assume that the cause of sequence decay remains unchanged: the single exponential decay, and SPAM errors captured in parameters $A$ and $B$, must be consistent. We experimentally find that behavior remains consistent, by comparing standard RB before and after optimization (Fig. 2 and Fig. 3). In addition, leakage into the larger Hilbert space outside of the qubit subspace is assumed to penalize sequence fidelity [25]. The results show that small leakage errors to higher levels penalize fidelity for single-qubit gates (Fig. 1b), and can be minimized for two-qubit gates, (Fig. 2). We underline that the reference and interleaved RB data should be verified for self-consistency [5, 16]. As a final note, randomized benchmarking and hence ORBIT work best when a gate has sufficient fidelity to construct a decay curve.

We have experimentally tested a new approach for optimizing quantum control using randomized benchmarking. This has been shown to be effective for improving single- and two-qubit gates, minimizing gate bleedthrough, and identifying control crosstalk. These experiments are a representative set of control problems for realizing high fidelity gates on large quantum systems. We believe ORBIT can be a generic tool for implementing closed-loop optimization in experimental systems, due to its speed, accuracy and platform independence.

We thank F. Wilhelm and D. Egger for helpful discussions on gate optimization and the Nelder-Mead algorithm. We also thank A. N. Korotkov and A. Veitia for help in implementing Clifford-based RB. This work was supported by the Office of the Director of National Intelligence (ODNI), Intelligence Advanced Research Projects Activity (IARPA), through the Army Research Office grants W911NF-09-1-0375 and W911NF-10-1-0334. All statements of fact, opinion or conclusions contained herein are those of the authors and should not be construed as representing the official views or policies of IARPA, the ODNI, or the U.S. Government. Devices were made at the UC Santa Barbara Nanofabrication Facility, a part of the NSF-funded National Nanotechnology Infrastructure Network, and at the NanoStructures Cleanroom Facility. J. Kelly and R. Barends contributed equally to this work.

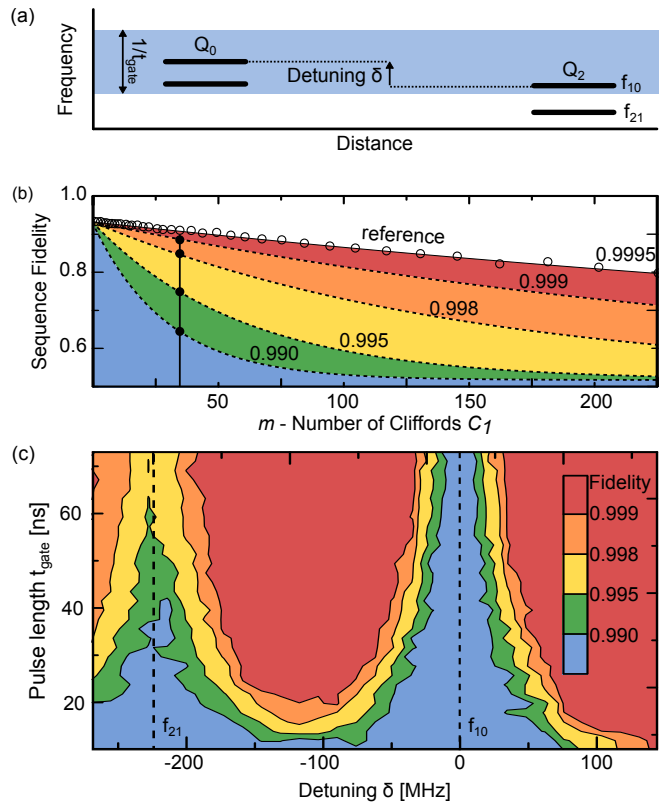

FIG. 4. (Color online) Mapping control crosstalk. (a) Energy level diagram of the qubits $Q_2$ and $Q_0$. Control pulses are applied to the XY lines of $Q_2$ and $Q_0$; the latter is swept in detuning $\delta$ and gate length $t_\text{gate}$. Control crosstalk can be exacerbated by small detunings or fast gates. (b) Single qubit benchmarking of $Q_2$. The colored regions indicate different ranges of reference fidelity. The vertical cut indicates the $m$ value used to discriminate between regions. (c) The inferred gate fidelity ($m = 35$, $k = 20$) versus detuning and gate length [16].

---

* These authors contributed equally to this work
[1] M. Shapiro and P. Brumer, Principles of the Quantum Control of Molecular Processes, by Moshe Shapiro, Paul Brumer, pp. 250. ISBN 0-471-24184-9. Wiley-VCH, February 2003. **1**

[2] M. A. Nielsen and I. L. Chuang, *Quantum computation and quantum information* (Cambridge university press, 2010).
[3] A. G. Fowler, M. Mariantoni, J. M. Martinis, and A. N. Cleland, Phys. Rev. A **86**, 032324 (2012).
[4] J. Benhelm, G. Kirchmair, C. F. Roos, and R. Blatt, Nature Physics **4**, 463 (2008).
[5] R. Barends, J. Kelly, A. Megrant, A. Veitia, D. Sank, E. Jeffrey, T. C. White, J. Mutus, A. G. Fowler, B. Campbell, et al., arXiv:1402.4848 (2014).
[6] K. Brown, A. Wilson, Y. Colombe, C. Ospelkaus, A. Meier, E. Knill, D. Leibfried, and D. Wineland, Phys. Rev. A **84**, 030303 (2011).
[7] J. L. O'Brien, G. Pryde, A. Gilchrist, D. James, N. Langford, T. Ralph, and A. White, Phys. Rev. Lett. **93**, 080502 (2004).
[8] T. Yamamoto, M. Neeley, E. Lucero, R. Bialczak, J. Kelly, M. Lenander, M. Mariantoni, A. OConnell, D. Sank, H. Wang, et al., Phys. Rev. B **82**, 184515 (2010).
[9] E. Magesan, J. Gambetta, and J. Emerson, Phys. Rev. Lett. **106**, 180504 (2011).
[10] A. Córcoles, J. M. Gambetta, J. M. Chow, J. A. Smolin, M. Ware, J. Strand, B. Plourde, and M. Steffen, Phys. Rev. A **87**, 030301 (2013).
[11] S. Gustavsson, O. Zwier, J. Bylander, F. Yan, F. Yoshihara, Y. Nakamura, T. P. Orlando, and W. D. Oliver, Phys. Rev. Lett. **110**, 040502 (2013).
[12] R. Barends, J. Kelly, A. Megrant, D. Sank, E. Jeffrey, Y. Chen, Y. Yin, B. Chiaro, J. Mutus, C. Neill, et al., Phys. Rev. Lett. **111**, 080502 (2013).
[13] F. Motzoi, J. Gambetta, P. Rebentrost, and F. K. Wilhelm, Phys. Rev. Lett. **103**, 110501 (2009).
[14] E. Lucero, J. Kelly, R. C. Bialczak, M. Lenander, M. Mariantoni, M. Neeley, A. OConnell, D. Sank, H. Wang, M. Weides, et al., Phys. Rev. A **82**, 042339 (2010).
[15] E. Magesan, J. M. Gambetta, B. Johnson, C. A. Ryan, J. M. Chow, S. T. Merkel, M. P. da Silva, G. A. Keefe, M. B. Rothwell, T. A. Ohki, et al., Phys. Rev. Lett. **109**, 080505 (2012).
[16] See supplementary information for discussion of ORBIT scaling with fidelity and additional data for Figures 3 and 4.
[17] A. G. Fowler, arXiv:1401.2466 (2014).
[18] J. M. Martinis and M. R. Geller, arXiv:1402.5467 (2014).
[19] N. Khaneja, T. Reiss, C. Kehlet, T. Schulte-Herbrüggen, and S. J. Glaser, Journal of Magnetic Resonance **172**, 296 (2005).
[20] D. Egger and F. Wilhelm, Superconductor Science and Technology **27**, 014001 (2014).
[21] T. Caneva, M. Murphy, T. Calarco, R. Fazio, S. Montangero, V. Giovannetti, and G. E. Santoro, Phys. Rev. Lett. **103**, 240501 (2009).
[22] D. Egger and F. K. Wilhelm, Submitted (2014).
[23] J. M. Gambetta, A. Córcoles, S. Merkel, B. Johnson, J. A. Smolin, J. M. Chow, C. A. Ryan, C. Rigetti, S. Poletto, T. A. Ohki, et al., Phys. Rev. Lett. **109**, 240504 (2012).
[24] R. Schutjens, F. A. Dagga, D. Egger, and F. Wilhelm, Phys. Rev. A **88**, 052330 (2013).
[25] J. M. Epstein, A. W. Cross, E. Magesan, and J. M. Gambetta, arXiv:1308.2928 (2013).

# Supplementary Information for 'Optimal quantum control using randomized benchmarking'

J. Kelly,[1, *] R. Barends,[1, *] B. Campbell,[1] Y. Chen,[1] Z. Chen,[1] B. Chiaro,[1] A. Dunsworth,[1] A. G. Fowler,[1,2] I. Hoi,[1] E. Jeffrey,[1] A. Megrant,[1] J. Mutus,[1] C. Neill,[1] P.J.J. O'Malley,[1] C. Quintana,[1] P. Roushan,[1] D. Sank,[1] A. Vainsencher,[1] J. Wenner,[1] T. C. White,[1] A. N. Cleland,[1] and John M. Martinis[1]

[1]*Department of Physics, University of California, Santa Barbara, CA 93106, USA*
[2]*Centre for Quantum Computation and Communication Technology,
School of Physics, The University of Melbourne, Victoria 3010, Australia*

## SCALING OF THE SENSITIVITY OF ORBIT WITH ERROR

Here, we derive the sensitivity of the sequence fidelity to gate error, and show that the sensitivity to fractional error is constant – hence ORBIT can in principle scale to arbitrarily small errors.

The sequence fidelity decays with $m$ following $F = Ap^m + B$. For the single-qubit case: $p = 1 - 2r$, with $r$ the error per Clifford. The variation in sequence fidelity with gate error is then $dF/dr = -2Am(1-2r)^{m-1}$. The optimal value of $m$ to operate ORBIT is at the characteristic decay of the sequence fidelity $m' = -1/\ln(1-2r)$ [this becomes clear when expressing the sequence fidelity as $F = A\exp(-m/m') + B$]. To quantify the scaling of the sensitivity with gate error, we evaluate the sensitivity at $m'$:

$$\left.\frac{dF}{dr}\right|_{m=m'} = \frac{2A}{e(1-2r)\ln(1-2r)} \approx -\frac{A}{er}, \quad (S1)$$

with the right side when expanding for small $r$, keeping the lowest order term.

Importantly, the sensitivity to fractional error $(dr/r)$ is constant,

$$S = \frac{dF}{dr/r} = \left.r\frac{dF}{dr}\right|_{m=m'} \approx -\frac{A}{e}. \quad (S2)$$

This is a crucial result, as it implies that ORBIT scales to arbitrarily small error: the sensitivity is the same when improving the fidelity of a 99.0% gate to 99.9%, or a 99.99% gate to 99.999%; only the choice for $m$ is different.

As an example, Eq. S1 is plotted in Fig. S1 for two cases: $r = 0.001$ and $0.0005$ ($A = 0.5$). These cases reach a maximum sensitivity at $m' = 500$ and $m' = 1000$ respectively. When halving the error the optimal $m$ and sensitivity double, as expected. We note that the sensitivity is retained for a wide range of $m$ around the optimum, therefore the choice of $m$ need not be exact. This is useful for improving gates, as we generally operate at a fixed $m$, and changes in $r$ will affect $m'$.

## CZ GATE FIDELITY BEFORE AND AFTER NELDER-MEAD OPTIMIZATION

The reference and interleaved randomized benchmarking data for the CZ gate, for Fig. 2 in the main text, are shown in Fig. S2. Figure S2a is before improvement, Fig. S2b after.

As a self-consistency check, we can calculate the expected error per Clifford using the derivation in Ref. [1]. We assume that gate errors are small and uncorrelated, such that adding errors is a good approximation. The expected error per Clifford is $r_{\text{ref,expected}} = 8.25 r_{\text{SQ}} + 1.5 r_{\text{CZ}}$ with $r_{\text{SQ}}$ the average single-qubit gate error and $r_{\text{CZ}}$ the CZ gate error. Assuming $r_{\text{SQ}} = 0.001$, we compute $r_{\text{ref,expected,before}} = 0.0318$ and $r_{\text{ref,expected,after}} = 0.0185$ which are close to the experimental values of $r_{\text{ref,before}} = 0.0361$ and $r_{\text{ref,after}} = 0.0188$.

## CONTROL CROSSTALK DATA

The sequence fidelity data, for Fig. 4 in the main text, are shown in Fig. S3.

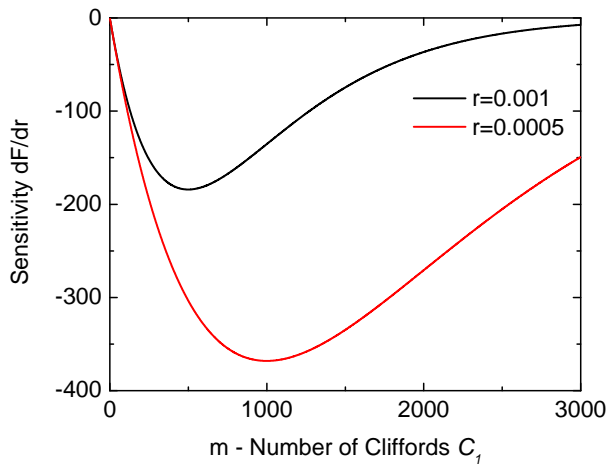

FIG. S1. The sensitivity of sequence fidelity to gate error (Eq. S1) for error per Clifford $r = 0.001$ and $r = 0.0005$.

[*] These authors contributed equally to this work
[1] R. Barends, J. Kelly, A. Megrant, A. Veitia, D. Sank, E. Jeffrey, T. C. White, J. Mutus, A. G. Fowler, B. Campbell, et al., arXiv:1402.4848 (2014).

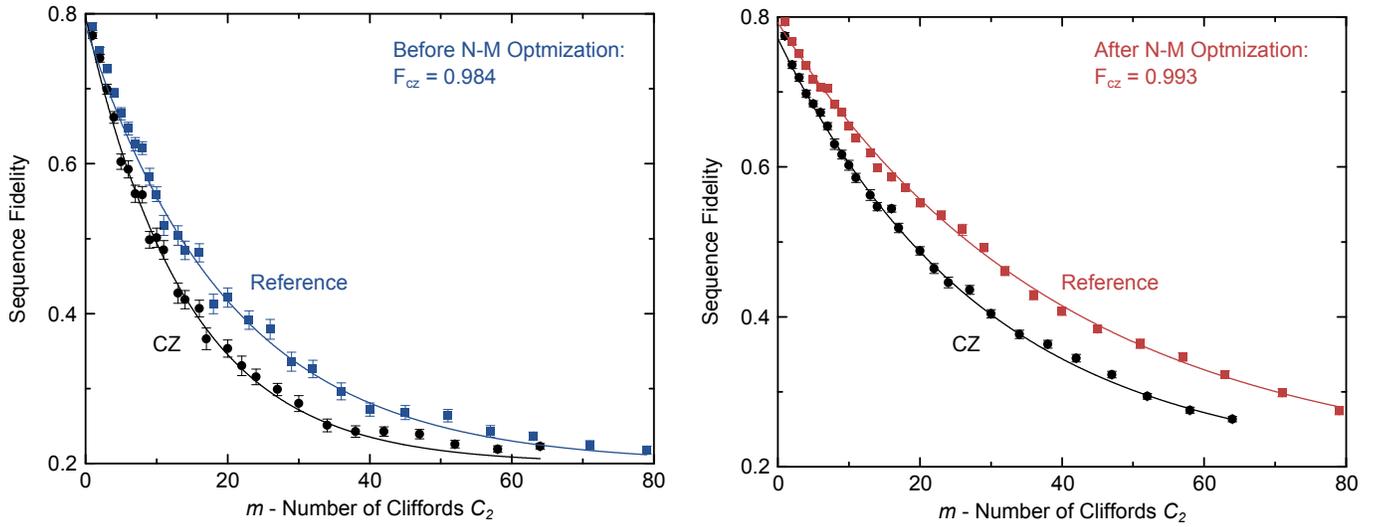

FIG. S2. Two-qubit randomized benchmarking data for Fig. 2 in the main text, showing the decay of the sequence fidelity of the reference and when interleaved with the CZ gate ($k = 50$). (a) Before Nelder-Mead optimization. Reference error: $r_{\text{ref}} = 0.0361$, interleaved error: $r_{\text{ref}+\text{CZ}} = 0.0511$, extracted CZ error: $r_{\text{CZ}} = 0.0157$. (b) After Nelder-Mead optimization. Reference error: $r_{\text{ref}} = 0.0188$, interleaved error: $r_{\text{ref}+\text{CZ}} = 0.0254$, extracted CZ error: $r_{\text{CZ}} = 0.0068$.

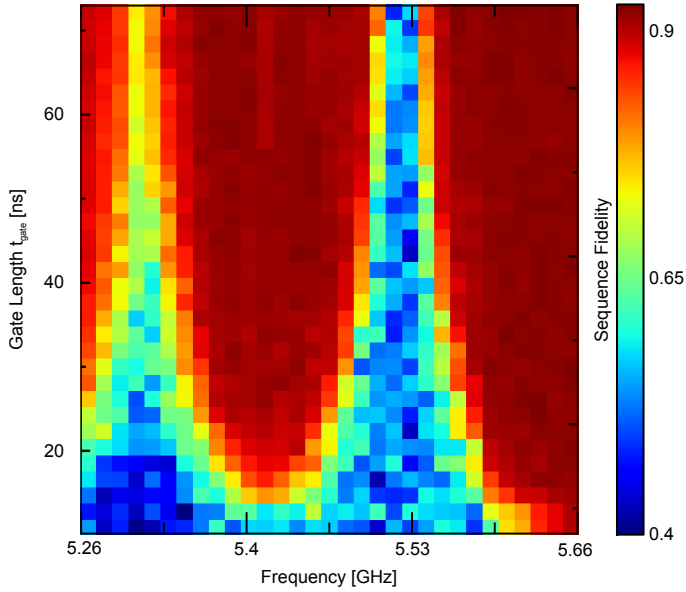

FIG. S3. Sequence fidelity data for Fig. 4 in the main text at $m = 35$ ($k = 20$).